\documentclass[9pt,conference]{IEEEtran}
\IEEEoverridecommandlockouts

\ifCLASSINFOpdf
   \usepackage[pdftex]{graphicx}
\else
\fi

\usepackage{amsmath}
\usepackage{algorithm,algpseudocode,amssymb,cite}

\begin{document}
%
\title{Joint recovery algorithms using difference of innovations for distributed compressed sensing}

\author{\IEEEauthorblockN{Diego Valsesia}
\IEEEauthorblockA{Politecnico di Torino\\
Email: diego.valsesia@polito.it}
\and
\IEEEauthorblockN{Giulio Coluccia}
\IEEEauthorblockA{Politecnico di Torino\\
Email: giulio.coluccia@polito.it}
\and
\IEEEauthorblockN{Enrico Magli}
\IEEEauthorblockA{Politecnico di Torino\\
Email: enrico.magli@polito.it}

\thanks{This work is supported by the European Research Council under the European Community’s Seventh Framework Programme (FP7/2007-2013) / ERC Grant agreement n.279848.}
}


%


\maketitle

\begin{abstract}
Distributed compressed sensing is concerned with representing an ensemble of jointly sparse signals using as few linear measurements as possible. Two novel joint reconstruction algorithms for distributed
compressed sensing are presented in this paper. These algorithms are based on the idea of using one of the signals as side information; this allows to exploit joint sparsity in a more effective way with respect to existing schemes. They provide gains in reconstruction quality, especially when the nodes acquire few measurements, so that the system is able to operate with fewer measurements than is required by other existing schemes. We show that the algorithms achieve better performance with respect to the state-of-the-art.
\end{abstract}

\vspace*{0.2cm}
\section{INTRODUCTION}
\label{sec:intro}
\vspace*{0.2cm}
Distributed compressed sensing (DCS) has recently attracted great interest as an efficient technique for acquiring data in a distributed fashion \cite{UniversalProjectionsDCS,distributedCS,DCS_InfoTheory}. DCS relies on the theory of compressed sensing (CS) to reduce the dimensionality of the signal acquired by each node of the distributed network, supposed to be sparse under some basis, by means of random projections. On the other hand, it also exploits the inter-correlation among the different signals in the ensemble to lower the number of measurements that each node needs to acquire, without requiring cooperation among nodes. Sensor networks represent the foremost application that can benefit from this technique because of their need of simple signal representations while meeting strict low complexity constraints (\emph{e.g.}, see \cite{NowakSensor}). 
A joint reconstruction algorithm can outperform separate recovery of the sensors' signals exploiting the correlation among them. Joint sparsity models have been proposed to account for sparsity patterns appearing in an ensemble of signals, e.g. signals acquired by nodes of a sensor network. According to the correlation model different algorithms have been proposed, such as the \emph{Texas Hold 'Em} \cite{texas}, and the \emph{Sort} and \emph{Intersect} algorithms \cite{Coluccia} for the JSM-1 model, in which signals are composed of a sparse common component plus a sparse innovation component. Concerning other models, we can find the DCS-SOMP \cite{distributedCS} method for JSM-2 (common sparse supports) and the TECC and ACIE algorithms for JSM-3 \cite{distributedCS} (nonsparse common component, sparse innovations).

This paper proposes two novel joint reconstruction algorithms. The algorithms are based on the idea of using one signal in the ensemble as side information. This allows to devise recovery schemes that attempt to reconstruct the difference between a signal and the side information, rather than an individual signal. In particular, we show that the proposed algorithms obtain lower reconstruction error with respect to other existing algorithms \cite{texas}\cite{Coluccia}, and minimize the number of measurements that have to be collected by each node.  

\vspace*{0.2cm}
\section{BACKGROUND}
\label{sec:bkg}
\vspace*{0.2cm}
\subsection{Compressed sensing}
Compressed sensing is a novel theory for signal sampling and acquisition \cite{CandesIntro,CS_donoho}. It is able to acquire signals in an already compressed fashion, i.e., using fewer coefficients than is dictated by the classical Nyquist-Shannon theory. Let us consider a signal $\mathbf{x} \in \mathbb{R}^n$, having a sparse representation under basis $\Psi \in \mathbb{R}^{n \times n}$:
$\mathbf{x} = \Psi\boldsymbol{\theta} \hspace{0.2cm} \mathrm{with} \hspace{0.2cm} \left\Vert \boldsymbol{\theta} \right\Vert_0 = k \ll n$, being $\left\Vert \boldsymbol{\theta} \right\Vert_0$ the $l_0$ norm of $\boldsymbol{\theta}$, \emph{i.e.}, the number of its nonzero entries. 
We acquire measurements as a vector of random projections $\mathbf{y} = \Phi\mathbf{x} = \Phi\Psi\boldsymbol{\theta}$, $\mathbf{y} \in \mathbb{R}^m$, using a sensing matrix $\Phi \in \mathbb{R}^{m \times n}$.
The best way to recover the original signal from its measurements is by solving an optimization problem trying to minimize the $l_0$ norm of the signal in the sparsity domain.
However, this problem is computationally intractable due to its NP-hard complexity, so it is common to consider a relaxed form using the $l_1$ norm, which can be solved by means of linear programming techniques:
\begin{equation*}
\hat{\boldsymbol{\theta}} = \arg \underset{\boldsymbol{\theta}}{\text{min}} \left\Vert \boldsymbol{\theta} \right\Vert_1 \hspace{.2cm} \text{subject to} \hspace{.2cm} \mathbf{y} = \Phi\Psi\boldsymbol{\theta} .
\end{equation*}   
This method is equivalent to $l_0$ norm minimization provided that the sensing matrix satisfies the Restricted Isometry Property (RIP) with constant $\delta_{2k}<\sqrt{2}-1$ \cite{CandesRIP}. The number of measurements to be acquired is typically $m = O\left( k\log \frac{n}{k} \right)$. A quadratically-constrained variant of the previous optimization problem is often used when dealing with noise with norm bounded by $\varepsilon$ affecting the measurements: 
\begin{align}
\label{BPDN}
\hat{\boldsymbol{\theta}} = \arg \underset{\boldsymbol{\theta}}{\text{min}} \left\Vert \boldsymbol{\theta} \right\Vert_1 \hspace{.2cm} \text{subject to} \hspace{.2cm} \left\Vert \Phi\Psi\boldsymbol{\theta} -\mathbf{y} \right\Vert_2 \leq \varepsilon .
\end{align}

\vspace*{0.1cm}
\subsection{Distributed compressed sensing}
\vspace*{0.1cm}
\label{sec:dcs}
In a distributed scenario, an ensemble of signals with both intra- and inter-sensor correlations is considered. The notion of joint sparsity has been introduced in \cite{distributedCS} for the framework of DCS. Among the joint sparsity models discussed in \cite{distributedCS}, we focus on the JSM-1 and JSM-3 models, according to which the $J$ signals in the ensemble have sparse innovation components and sparse or nonsparse common component, respectively.
\begin{equation*}
\boldsymbol{\theta}_j = \boldsymbol{\theta}_C + \boldsymbol{\theta}_{I,j} \hspace{.2cm} \forall j \in \left[1,J\right]
\end{equation*}
\begin{equation*}
\left\Vert \boldsymbol{\theta}_C \right\Vert_0 = k_C  \hspace{.2cm} \text{and}  \hspace{.2cm} \left\Vert \boldsymbol{\theta}_{I,j} \right\Vert_0 = k_{I,j} \hspace{.2cm} \forall j \in \left[1,J\right]
\end{equation*}
A joint reconstruction algorithm can leverage the structure of the joint sparsity model to improve performance, namely to achieve higher quality for the same number of measurements or decrease the number of measurements needed to achieve the same quality. 

Some of the existing joint reconstruction algorithms for the JSM-1 model include the \emph{weighted} $l_1$ \emph{minimization} proposed in \cite{distributedCS}, which requires numerical optimization of the weights, the \emph{Texas Hold 'Em} \cite{texas} and the \emph{Sort} and \emph{Intersect} algorithms \cite{Coluccia}. \emph{Texas Hold 'Em} averages a subset of the measurements of all the sensors to estimate the common component, which is then subtracted from the measurements to recover the innovations. \emph{Sort} and \emph{Intersect} also estimate the common component, in a nonlinear way. In particular, in \emph{Sort} the coefficients of a first estimate of the signal are sorted by decreasing magnitude and compared to ones of the side information to decide which positions belong to the common support, while in \emph{Intersect} the supports of a first estimate of the signal and of the side information are intersected to find the support of the common component. 

As far as the JSM-3 model is concerned, the common component is not sparse, hence a joint recovery algorithm must be used in order to acquire fewer than $n$ measurements per signal. The Transpose Estimation of Common Component (TECC) algorithm described in \cite{distributedCS} estimates the common component by stacking all the measurements from all nodes in a single problem and then recover the innovations alone by cancelling the measurements of the estimated common component. A key requirement for TECC to work is having different sensing matrices $\Phi_j$ so that, when stacked, they span the whole $\mathbb{R}^n$. The Alternating Common and Innovation Estimation (ACIE) algorithm \cite{distributedCS} is an improvement over TECC, based on an iterative scheme that alternates improvements on the estimate of the common component with improvements in the estimate of the innovations, at the expense of a great computational complexity. As in TECC, each node must have a different sensing matrix in order to work properly. This is different from the scenario we are considering, in which all nodes use the same sensing matrix.

\vspace*{0.2cm}
\section{PROPOSED ALGORITHMS}
\label{proposed}
\vspace*{0.2cm}
The proposed algorithms focus on the JSM-1 and JSM-3 models discussed in section \ref{sec:dcs}. We also suppose that side information is available at the decoder in the form of perfect knowledge of one of the signals. From now on we suppose, without loss of generality, that the known signal is $\mathbf{x}_1$. From a practical perspective, the requirement of side information is not a limitation; under the JSM-1 model the SI signal is sparse, so we can acquire $m_1<n$ measurements, with $m_1$ large enough to get the desired accuracy. Under the JSM-3 model the SI signal is not sparse and has to be acquired uncompressed or compressed using a standard technique. However, as the number of nodes increases, the overhead due to side information becomes negligible. The savings in the number of measurements to be acquired by the other nodes outweigh the small overhead due to the acquisition of side information, thus making the framework interesting even for the JSM-3 model, where the SI signal is not compressed. 

\begin{figure*}

\begin{minipage}[b]{.48\linewidth}
  \centering
  \centerline{\includegraphics[width=1.0\linewidth]{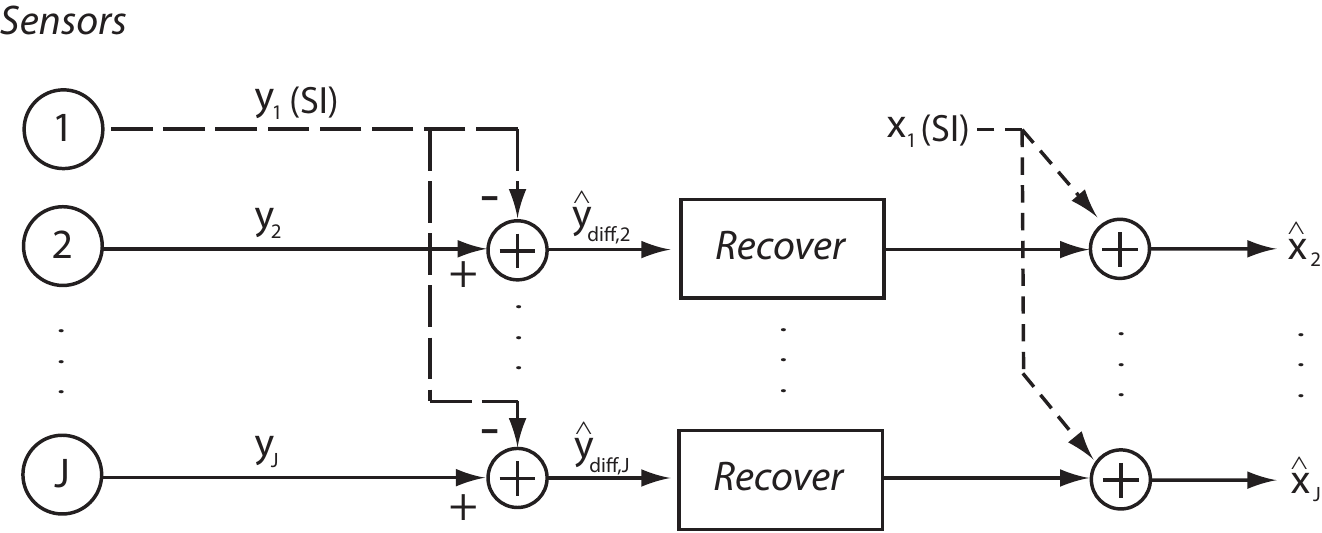}}
  \centerline{(a) \emph{DOI} algorithm}\medskip
\end{minipage}
\hfill
\begin{minipage}[b]{0.48\linewidth}
  \centering
  \centerline{\includegraphics[width=1.0\linewidth]{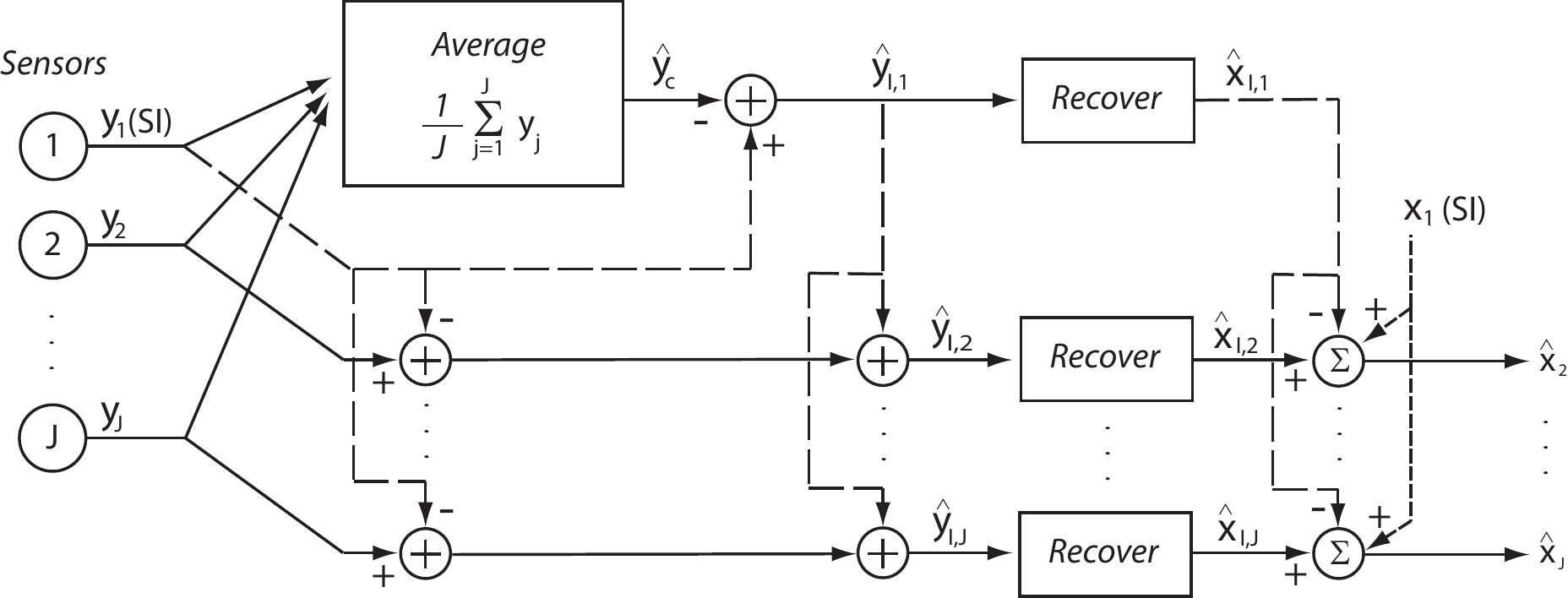}}
  \centerline{(b) \emph{Texas DOI} algorithm}\medskip
\end{minipage}
\caption{Joint reconstruction algorithms}
\vspace*{-0.3cm}
\label{alg_schem}

\end{figure*}

\vspace*{0.1cm}
\subsection{Difference-Of-Innovations (DOI) algorithm}
\vspace*{0.1cm}
\begin{algorithm}
\label{diff_algo}
\caption{\emph{DOI} algorithm}
\begin{algorithmic}
\Require $A=\Phi\Psi$
\For {$j$ in $2:J$}
\State Compute $\mathbf{y}_{\text{diff,$j$}}=\mathbf{y}_j-\mathbf{y}_1$
\State Recover $\boldsymbol{\theta}_{\text{diff,$j$}}$ from $\mathbf{y}_{\text{diff,$j$}}$
\State $\hat{\boldsymbol{\theta}}_j=\boldsymbol{\theta}_1 + \boldsymbol{\theta}_{\text{diff,$j$}}$
\EndFor
\end{algorithmic}
\end{algorithm}
The underlying idea of this first algorithm is to exploit side information to eliminate the need to recover the common component. Figure \ref{alg_schem}(a) presents a schematic representation of the algorithm, also outlined in Alg.1. Proceeding pairwise by using the side information and each of the $J$ signals in the ensemble, it is possible to compute the difference between the measurements of the side information and those of signal $j$. This removes exactly any component that is common to the two signals, so we are left with measurements of the difference of the innovation components. It is then possible to recover the difference signal from these measurements using any recovery procedure such as $l_1$ minimization. Once the difference signal is recovered, it is then sufficient to add the side information to fully recover signal $j$. It is interesting to notice that, unlike \cite{texas}, this algorithm does not introduce any error due to an inexact estimation of the common component. However, the difference signal is, in general, less sparse than the individual innovation component, so we can expect performance gains only when innovations are significantly sparser than the common component. In particular, as a rule of thumb, we expect gains for $k_C \geq 2k_{I,j}$. This condition is easily satisfied for highly correlated ensembles where there is a dominant common component and much sparser innovations.
Also note that the \emph{DOI} algorithm can be readily implemented in a parallel manner since any recovery only involves the side information and the measurements of the signal to be recovered.

\vspace*{0.1cm}
\subsection{Performance bound}
\vspace*{0.1cm}
\label{diff_pb}
In this section we evaluate a bound to the reconstruction error of the \emph{DOI} algorithm. In particular, it can be shown that:
\begin{align}
\left\Vert \hat{\boldsymbol{\theta}}_{j}-\boldsymbol{\theta}_{j}\right\Vert_2 
& = \left\Vert \boldsymbol{\theta}_{I,1}-\boldsymbol{\theta}_{I,j}+\boldsymbol{\theta}_{\text{diff,$j$}}\right\Vert_2 \notag \\
 & = \left\Vert \boldsymbol{\theta}_{\text{diff,$j$}}-\left(\boldsymbol{\theta}_{I,j}-\boldsymbol{\theta}_{I,1}\right)\right\Vert_2  \leq  C\varepsilon
\end{align}
where $\varepsilon = \left\Vert \mathbf{y}_\text{diff,$j$} - \Phi\left(\mathbf{x}_{I,1} - \mathbf{x}_{I,j} \right)\right\Vert_2$ is the norm of the noise affecting the measurements of the difference signal and $C$ is a constant that depends on the method used for reconstruction and on the RIP constant of the sensing matrix. When there is no quantization, or other sources of noise, we have $\varepsilon=0$  and, provided that there are enough measurements available, reconstruction is perfect. This means that the \emph{DOI} algorithm can achieve perfect reconstruction, unlike \cite{texas} which is limited by the residual noise in the averaging procedure.

\vspace*{0.1cm}
\subsection{\emph{Texas DOI} algorithm}
\vspace*{0.1cm}
\begin{algorithm}
\caption{\emph{Texas DOI} algorithm}
\begin{algorithmic}
\Require $J$,$A=\Phi\Psi$, $k_I$
\State $\hat{\mathbf{y}}_C = \frac{1}{J}\underset{j=1}{\overset{J}{\sum}}\mathbf{y}_{j} $
\State $\hat{\mathbf{y}}_{I,1} = \mathbf{y}_1 - \hat{\mathbf{y}}_C$
\State Recover $\hat{\boldsymbol{\theta}}_{I,1}$ from $\hat{\mathbf{y}}_{I,1}$
\For {$j$ in $2:J$}
\State $\mathbf{y}_{\text{diff,$j$}} = \mathbf{y}_j - \mathbf{y}_1$
\State $\hat{\mathbf{y}}_{I,j} = \mathbf{y}_{\text{diff,$j$}} + \hat{\mathbf{y}}_{I,1}$
\State Recover $\hat{\boldsymbol{\theta}}_{I,j}$ from $\hat{\mathbf{y}}_{I,j}$
\State $\hat{\boldsymbol{\theta}}_j = \theta_1 - \hat{\boldsymbol{\theta}}_{I,1} + \hat{\boldsymbol{\theta}}_{I,j}$
\EndFor
\end{algorithmic}
\end{algorithm}

The second proposed algorithm, called \emph{Texas DOI}, attempts at overcoming the drawbacks of \emph{DOI} and \cite{texas}. Figure \ref{alg_schem}(b) presents a schematic representation of the algorithm, also outlined in Alg.2. Albeit it maintains the idea of \cite{texas} of averaging a fraction of the collected measurements, it avoids any explicit reconstruction of the common component, but rather employs the measurements of the common component, and combines this with the use of side information in a fashion similar to the \emph{DOI} algorithm. In particular, side information is used to get differences of innovation components, but the measurements of the innovation component of the side information ($y_{I,1}$) can be obtained subtracting the output of the averaging procedure from the original SI measurements. This allows to perform recovery on the innovation component of a single signal by removing $y_{I,1}$.   
The algorithm can be implemented in a parallel manner if the averaging procedure and the recovery of $x_{I,1}$ are computed first. In fact all the remaining operations for the recovery of the signals only involve the available quantities related to the side information and the measurements of the signal to be recovered.

\vspace*{0.1cm}
\subsection{Performance bound}
\vspace*{0.1cm}
In this section we show a bound to the reconstruction error of the \emph{Texas DOI} algorithm, whose proof is reported in the appendix. In particular, it can be shown that:
\begin{align}
\label{pbTexasFinal}
\left\Vert \hat{\boldsymbol{\theta}}_{j}-\boldsymbol{\theta}_{j}\right\Vert_2 \leq 2C\cdot\frac{\sqrt{1+\delta_k}}{\sqrt{J}}\eta ,
\end{align}
being $\delta_k$ the RIP constant of matrix $A=\Phi\Psi$, and $\left\Vert \boldsymbol{\theta}_{I,j}\right\Vert_2 = \eta$ for all $j\in\left[1,J\right]$.
This analysis points out some interesting properties of the algorithm. Even if quantization or other sources of noise are not considered by the analysis, the algorithm is still affected by a certain reconstruction error, in the same way as \cite{texas}. This does not happen with \emph{DOI}, which we showed in section \ref{diff_pb} to be exact. Here, and in \cite{texas}, the limiting factor lies in the averaging procedure that imposes a floor on the reconstruction error, which cannot be overcome by adjustments on the rate. However, this error floor decreases as $\frac{1}{\sqrt{J}}$, so the performance of an ensemble with a large number of signals may indeed be limited by the quantization rate or other sources of noise rather than the averaging procedure.

\vspace*{0.2cm}
\section{EXPERIMENTAL RESULTS}
\vspace*{0.2cm}

\begin{figure}[t]
\begin{minipage}[t]{1.0\linewidth}
  \centering
  \centerline{\includegraphics[width=0.85\linewidth]{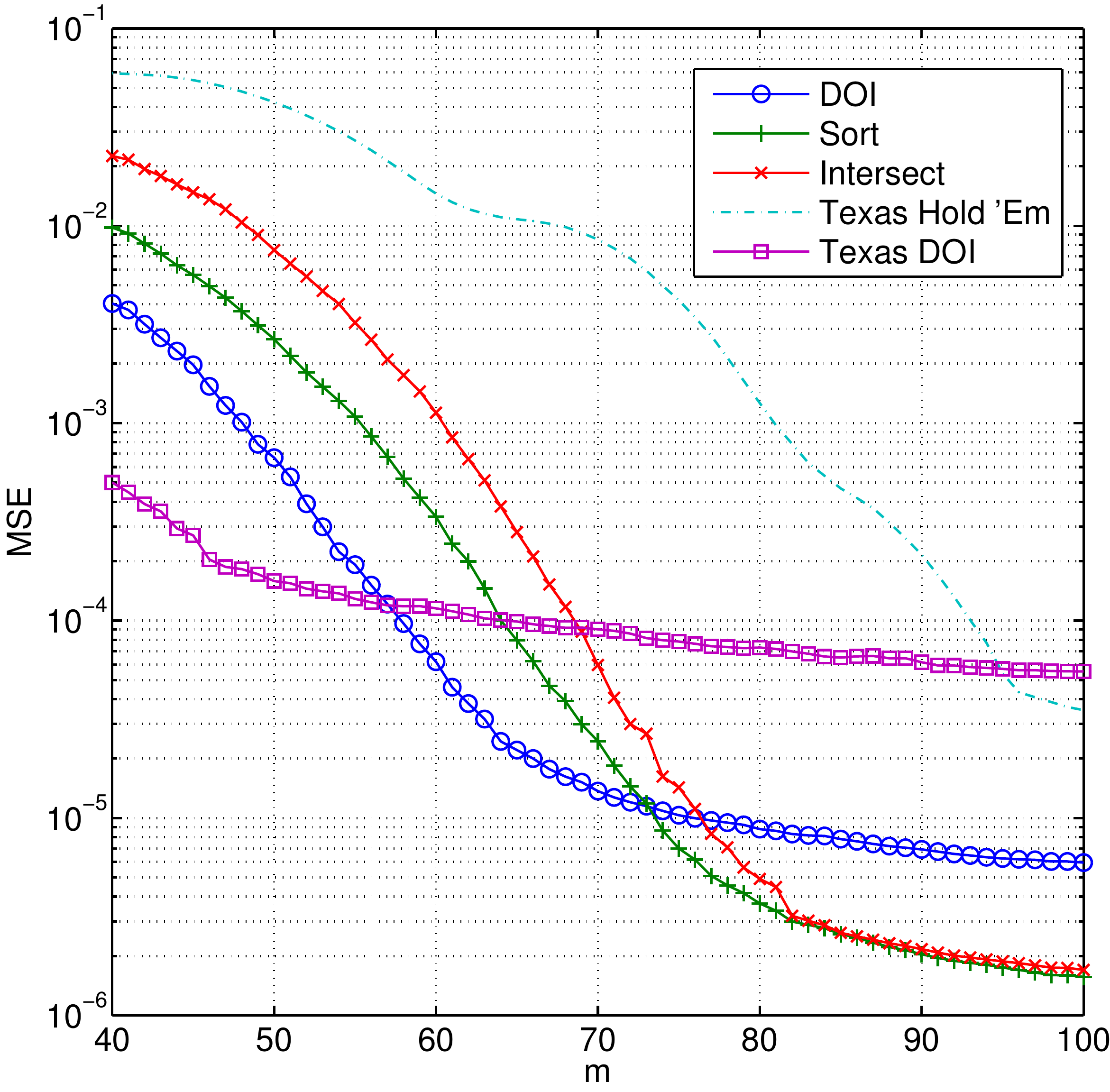}}
  \caption{JSM-1. Mean square error vs. number of measurements. $J=100$, $k_C=20$, $k_I=5$, $n=256$, $R=8bps$}\medskip
\end{minipage}
\label{err_vs_m}
\vspace*{-0.5cm}
\end{figure}

We compared the two proposed algorithms with some of the existing joint reconstruction algorithms in the literature for the JSM-1 model such as \emph{Texas Hold 'Em}\cite{texas}, \emph{Intersection} and \emph{Sort}\cite{Coluccia}.
The simulations have been performed using a sensing matrix with i.i.d. zero-mean Gaussian entries with unit-norm columns and a JSM-1 ensemble of signals sparse in the identity basis. The amplitude of the nonzero entries of the signal is randomly generated according to a standard Gaussian distribution. Each measurement is quantized using $R$ bits. For the \emph{Texas Hold 'Em} algorithm all the measurements of each node are considered community measurements, thus contributing to the estimation of the common component. Figure \ref{err_vs_m} shows that the proposed algorithms achieve lower MSE when few measurements are available. \emph{Texas DOI} can achieve the best performance but is not able to improve when $m$ increases due to the error floor in \eqref{pbTexasFinal}.  The \emph{DOI} algorithm leverages the side information to remove the common component, but the recovery step is performed on the difference of measurements of the innovation components, hence it is outperformed by \emph{Texas DOI} for low $m$. In fact, \emph{Texas DOI} is able to run the compressed sensing recovery procedures only for the individual innovation components, whereas \emph{DOI} has to recover the difference between two innovation signals, which is typically less sparse. When few measurements are available the algorithms in \cite{Coluccia} may have trouble in recovering the common component because they rely on an initial estimate of the unknown signals. Hence the greatest gains are achieved when a limited number of measurements is available. The \emph{Texas DOI} algorithm borrows ideas from both the \emph{Texas Hold 'Em} strategy and the \emph{DOI} procedure. \emph{Texas DOI} inherits the averaging procedure from \emph{Texas Hold 'Em}, which is very efficient when the number of nodes is large. However exploiting side information allows to improve over \emph{Texas Hold 'Em} when the signals are highly correlated and few measurements are available because \emph{Texas Hold 'Em} may have difficulties in recovering the common component. This allows each node to work closer to the minimum number of measurements needed by CS reconstruction for successful recovery. 

It should be noted that \emph{Texas Hold 'Em} does not use side information. However, the overhead due to acquiring more measurements to recover the side information with the desired accuracy is negligible in our simulations. As an example, in the case of $J=100$, the side information can be recovered from $m_1 \cong 5k=125$ measurements quantized with rate $R_1 = 8$. This means that the total number of bits is $(J-1)mR+m_1R_1$ for \emph{DOI} and \emph{Texas DOI} and $Jm'R$ for \emph{Texas Hold 'Em}, with $m' > m$ in order to achieve $Jm'R=(J-1)mR+m_1R_1$. However, typically $m'-m < 1$, so no extra measurement has to be assigned to \emph{Texas Hold 'Em} to compensate for the overhead of SI, which is completely negligible.

\begin{figure}[t]
\begin{minipage}[t]{1.0\linewidth}
  \centering
  \centerline{\includegraphics[width=0.85\linewidth]{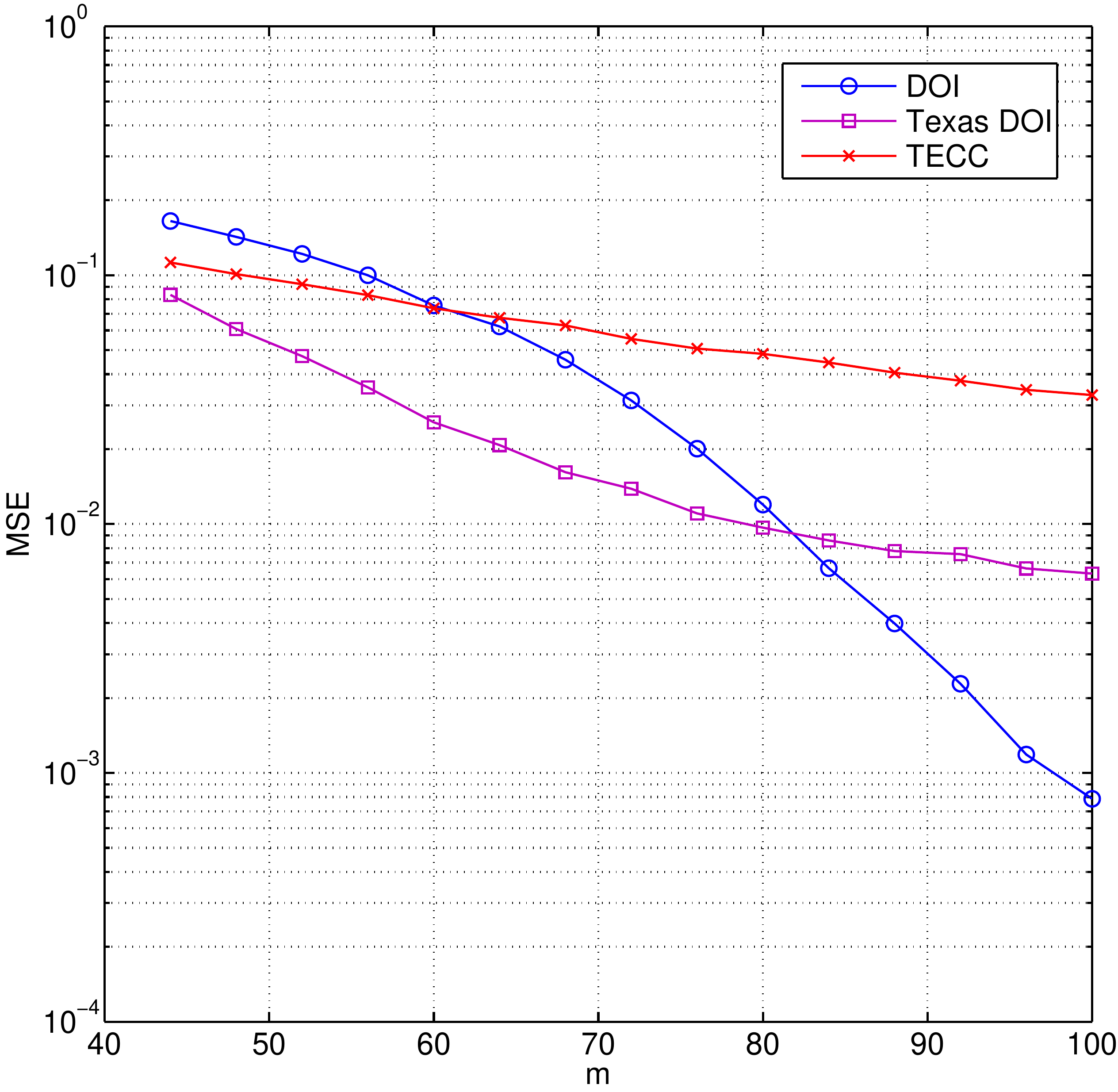}}
  \caption{JSM-3. Mean square error vs. number of measurements. $J=100$, $k_I=20$, $n=256$, $R=8bps$}\medskip
\end{minipage}
\label{jsm3}
\vspace*{-0.5cm}
\end{figure}

The algorithms have also been tested on the JSM-3 model of distributed compressed sensing and compared against the TECC algorithm presented in \cite{distributedCS}. As explained in section \ref{proposed} the proposed algorithms rely on the usage of the same sensing matrix for all nodes, while TECC requires different matrices. Figure \ref{jsm3} shows the MSE as a function of the number of measurements acquired by each node. The proposed algorithms are able to outperform TECC and confirm the behaviour presented for the JSM-1 model.

\vspace*{0.2cm}
\section{CONCLUSIONS}
\vspace*{0.2cm}
We proposed two novel joint reconstruction algorithms for the JSM-1 and JSM-3 models in distributed compressed sensing. Thanks to the use of side information, it is possible to devise methods that avoid the need to reconstruct the common component, thus allowing to deal with the case of a non-sparse common component in a straightforward manner. The algorithms provide performance gains over other existing techniques, especially when few measurements are available, thus allowing to decrease the number of measurements needed to achieve a target quality in the reconstruction or to improve quality for the same number of measurements.

\vspace*{-0.1cm}
\section{APPENDIX \\ Proof of the performance bound of \emph{Texas DOI}}
\vspace*{-0.3cm}

\begin{align}
\label{pbTexas}
\left\Vert \hat{\boldsymbol{\theta}}_{j}-\boldsymbol{\theta}_{j}\right\Vert_2  & = \left\Vert \left( \boldsymbol{\theta}_{1}-\hat{\boldsymbol{\theta}}_{I,1}+\hat{\boldsymbol{\theta}}_{I,j} \right)-\boldsymbol{\theta}_{j}\right\Vert_2 \notag \\ 
&=\left\Vert \boldsymbol{\theta}_{I,1}-\hat{\boldsymbol{\theta}}_{I,1}-\boldsymbol{\theta}_{I,j}+\hat{\boldsymbol{\theta}}_{I,j}\right\Vert_2 \notag \\
 & = \left\Vert \left(\boldsymbol{\theta}_{I,1}-\hat{\boldsymbol{\theta}}_{I,1}\right)+\left(\hat{\boldsymbol{\theta}}_{I,j}-\boldsymbol{\theta}_{I,j}\right)\right\Vert_2 \notag \\ 
 & \leq  \left\Vert \boldsymbol{\theta}_{I,1}-\hat{\boldsymbol{\theta}}_{I,1}\right\Vert_2 +\left\Vert \hat{\boldsymbol{\theta}}_{I,j}-\boldsymbol{\theta}_{I,j}\right\Vert_2 
\end{align}
Let us analyse how the innovation components are recovered.
\begin{align*}
\hat{\mathbf{y}}_{I,1} &= \mathbf{y}_1 - \hat{\mathbf{y}}_C = \mathbf{y}_{I,1} - \frac{1}{J}\underset{l=1}{\overset{J}{\sum}}\mathbf{y}_{I,l}
\end{align*}
\begin{align*}
\hat{\mathbf{y}}_{I,j} &= \mathbf{y}_j - \mathbf{y}_1 + \hat{\mathbf{y}}_{I,1} = \mathbf{y}_{I,j} - \frac{1}{J}\underset{l=1}{\overset{J}{\sum}}\mathbf{y}_{I,l}
\end{align*}
Let  $\mathbf{n} = \frac{1}{J}\underset{l=1}{\overset{J}{\sum}}\mathbf{y}_{I,l}$ denote the error in the estimation of the common component, due to imperfect cancellation of the innovations. Hence, $\hat{\mathbf{y}}_{I,j} = A\boldsymbol{\theta}_{I,j} -\mathbf{n}$ for $j\in\left[1,J\right]$. Suppose that we use a reconstruction procedure (e.g. BPDN \eqref{BPDN}) from noisy measurements that has a performance guarantee ensuring that the reconstruction error is proportional to the noise norm with a constant $C$ depending on the reconstruction method and on the RIP constant of $A$. Assuming that enough measurements have been acquired so that $A$ satisfies the RIP of order $k$ with constant $\delta_k$ with high probability, that $\left\Vert \boldsymbol{\theta}_{I,j}\right\Vert_2 = \eta$ for all $j\in\left[1,J\right]$ and that the $\boldsymbol{\theta}_{I,j}$'s are pairwise orthogonal (i.e. $\boldsymbol{\theta}_{I,j}^T\boldsymbol{\theta}_{I,k} = 0$ for $j\neq k$), we can write:
\begin{align}
\left\Vert \hat{\boldsymbol{\theta}}_{I,j}-\boldsymbol{\theta}_{I,j}\right\Vert_2 & \leq C\left\Vert \mathbf{n}\right\Vert_2 = C\left\Vert \frac{1}{J}\underset{j=1}{\overset{J}{\sum}}A \boldsymbol{\theta}_{I,j}\right\Vert_2 \notag \\ 
& \leq C\frac{\sqrt{1+\delta_k}}{\sqrt{J}}\left\Vert \boldsymbol{\theta}_{I,j}\right\Vert_2 =C\frac{\sqrt{1+\delta_k}}{\sqrt{J}}\eta
\end{align}
where $\delta_k$ is the RIP constant of matrix $A=\Phi\Psi$.
Finally, we can plug this result in \eqref{pbTexas} and we obtain:
\begin{align}
\left\Vert \hat{\boldsymbol{\theta}}_{j}-\boldsymbol{\theta}_{j}\right\Vert_2 & \leq \left\Vert \boldsymbol{\theta}_{I,1}-\hat{\boldsymbol{\theta}}_{I,1}\right\Vert_2 +\left\Vert \hat{\boldsymbol{\theta}}_{I,j}-\boldsymbol{\theta}_{I,j}\right\Vert_2 \notag \\ 
& \leq 2C\cdot\frac{\sqrt{1+\delta_k}}{\sqrt{J}}\eta
\end{align}

\begin{thebibliography}{1}

\bibitem{UniversalProjectionsDCS}
M.R. Duarte, M.B. Wakin, D.~Baron, and R.G. Baraniuk,
\newblock ``Universal distributed sensing via random projections,''
\newblock in {\em Information Processing in Sensor Networks, 2006. IPSN 2006.
  The Fifth International Conference on}, 0-0 2006, pp. 177 --185.

\bibitem{distributedCS}
D.~Baron, M.F. Duarte, M.B. Wakin, S.~Sarvotham, and R.G. Baraniuk,
\newblock ``Distributed compressive sensing,''
\newblock {\em Preprint}, Jan. 2009.

\bibitem{DCS_InfoTheory}
M.F. Duarte, M.B. Wakin, D.~Baron, S.~Sarvotham, and R.G. Baraniuk,
\newblock ``Measurement bounds for sparse signal ensembles via graphical
  models,''
\newblock {\em Information Theory, IEEE Transactions on}, vol. 59, no. 7, pp.
  4280--4289, 2013.

\bibitem{NowakSensor}
W.~Bajwa, J.~Haupt, A.~Sayeed, and R.~Nowak,
\newblock ``Compressive wireless sensing,''
\newblock in {\em Proceedings of the 5th international conference on
  Information processing in sensor networks}, New York, NY, USA, 2006, IPSN
  '06, pp. 134--142, ACM.

\bibitem{texas}
S.R. Schnelle, J.N. Laska, C.~Hegde, M.F. Duarte, M.A. Davenport, and R.G.
  Baraniuk,
\newblock ``Texas hold 'em algorithms for distributed compressive sensing,''
\newblock in {\em Acoustics Speech and Signal Processing (ICASSP), 2010 IEEE
  International Conference on}, march 2010, pp. 2886 --2889.

\bibitem{Coluccia}
G.~Coluccia, E.~Magli, A.~Roumy, and V.~Toto-Zarasoa,
\newblock ``Lossy compression of distributed sparse sources: a practical
  scheme,''
\newblock in {\em 2011 European Signal Processing Conference (EUSIPCO-2011)},
  Barcelona, Spain, Sept. 2011.

\bibitem{CandesIntro}
E.J. Candes and M.B. Wakin,
\newblock ``An introduction to compressive sampling,''
\newblock {\em IEEE Signal Processing Magazine}, vol. 25, no. 2, pp. 21--30,
  March 2008.

\bibitem{CS_donoho}
D.L. Donoho,
\newblock ``Compressed sensing,''
\newblock {\em Information Theory, IEEE Transactions on}, vol. 52, no. 4, pp.
  1289 --1306, april 2006.

\bibitem{CandesRIP}
E.J. Candes,
\newblock ``The restricted isometry property and its implications for
  compressed sensing,''
\newblock {\em Comptes Rendus Mathematique}, vol. 346, no. 9–10, pp. 589 --
  592, 2008.

\end{thebibliography}
\end{document}